\documentclass{CUP-JNL-DTM}%

\usepackage{graphicx}
\usepackage{multicol,multirow}
\usepackage{amsmath,amssymb,amsfonts}
\usepackage{mathrsfs}
\usepackage{amsthm}
\usepackage{rotating}
\usepackage{appendix}
\usepackage[natbib,style=apa]{biblatex}
\usepackage{ifpdf}
\usepackage[T1]{fontenc}
\usepackage{newtxtext}
\usepackage{newtxmath}
\usepackage{textcomp}
\usepackage{xcolor}
\usepackage{lipsum}
\usepackage[colorlinks,allcolors=blue]{hyperref}
\usepackage{lineno}

\theoremstyle{definition}

\numberwithin{equation}{section}
\usepackage{tablefootnote}

\jname{Data-Centric Engineering}
\articletype{Research article}
\jyear{2024}

\begin{document}

\begin{Frontmatter}

\title[Article Title]{Using Graph Neural Networks and Frequency Domain Data for Automated Operational Modal Analysis of Populations of Structures}

\author[1]{Xudong Jian}
\author[2]{Yutong Xia}
\author[3]{Gregory Duthé}
\author[1,3]{Kiran Bacsa}
\author[1,4]{Wei Liu}
\author[1,3]{Eleni Chatzi*}

\authormark{Xudong Jian \textit{et al}.}

\address[1]{\orgdiv{Future Resilient Systems}, \orgname{Singapore-ETH Centre}, \orgaddress{\city{Singapore}, \postcode{138602}, \country{Singapore}}}

\address[2]{\orgdiv{Institute of Data Science}, \orgname{National University of Singapore}, \orgaddress{\city{Singapore}, \postcode{117602}, \country{Singapore}}}

\address[3]{\orgdiv{Department of Civil, Environmental and Geomatic Engineering}, \orgname{ETH Zurich}, \orgaddress{\city{Zurich}, \postcode{8049},  \country{Switzerland}}. \email{chatzi@ibk.baug.ethz.ch}}

\address[4]{\orgdiv{Department of Industrial Systems Engineering and Management}, \orgname{National University of Singapore}, \orgaddress{\city{Singapore}, \postcode{117576}, \country{Singapore}}}

\authormark{Xudong Jian et al.}



\abstract{The Population-Based Structural Health Monitoring (\textit{PBSHM}) paradigm has recently emerged as a promising approach to enhance data-driven assessment of engineering structures by facilitating transfer learning between structures with some degree of similarity. In this work, we apply this concept to the automated modal identification of structural systems. We introduce a Graph Neural Network (GNN)-based deep learning scheme to identify modal properties, including natural frequencies, damping ratios, and mode shapes of engineering structures based on the Power Spectral Density (PSD) of spatially-sparse vibration measurements. Systematic numerical experiments are conducted to evaluate the proposed model, employing two distinct truss populations that possess similar topological characteristics but varying geometric (size, shape) and material (stiffness) properties. The results demonstrate that, once trained, the proposed GNN-based model can identify modal properties of unseen structures within the same structural population with good efficiency and acceptable accuracy, even in the presence of measurement noise and sparse measurement locations. The GNN-based model exhibits advantages over the classic Frequency Domain Decomposition (FDD) method in terms of identification speed, as well as against an alternate Multi-Layer Perceptron (MLP) architecture in terms of identification accuracy, rendering this a promising tool for \textit{PBSHM} purposes.}

\end{Frontmatter}

\section*{Impact Statement}
We have developed a Graph Neural Network (GNN)-based scheme for automated operational modal analysis. Numerical experiments demonstrate that the proposed model can efficiently and effectively identify natural frequencies, damping ratios, and mode shapes of different engineering structures that belong to a population. Notably, the GNN-based model outperforms the traditional Frequency Domain Decomposition method in terms of efficiency and surpasses a Multi-layer Perceptron in accuracy, positioning it as a promising tool for population-based structural health monitoring.


\section{Introduction}
Structural Health Monitoring (SHM) has evolved into a useful tool for engineering practice, showcasing its importance through various applications in the management and maintenance of engineering structures (\cite{farrar2012structural,an2019recent,kamariotis2022value}). Although further steps are needed for standardization of SHM schemes, which are often bespoke to the structure at hand, the overall practice has matured, with the primary methods for extracting valuable information from SHM data categorized into physics-based and data-driven approaches \citep{avendano2017gaussian,karakostas2024seismic}. The notable rise of data-driven approaches in recent years can be attributed not only to advancements in powerful Deep Learning (DL) tools, which favors computation, but also to the apparent robustness of schemes that capitalize on data against discrepancies between physical models and real-world structures \citep{cicirello2024physics,haywood2023discussing}. However, these approaches face challenges in interpretability and generalization, especially in civil engineering, where structures such as bridges are often uniquely designed to meet site-specific requirements \citep{lei2023interpretable}. Training data-driven models with large datasets that represent a variety of structures can enhance their generalization capabilities. However, data from real structures often lack, since SHM tends to so far mostly be applied onto specific case studies, of higher importance, due to budget and labor constraints. Consequently, the rich and diverse datasets necessary for training learning (data-driven) models are generally unavailable (\cite{sun2020review}). To address this challenge, new sensing technologies, such as mobile sensing schemes (\cite{jian2022indirect,stoura2023model,jian2024robotic}), are developed to gather more measurement data from a greater number of structures. Additionally, advanced analysis methods can be used to more efficiently extract deeper insights from the available data. Treating structures as a population of shared features could enable learning and knowledge transfer among group members. For instance, although bridges are often unique, they can be categorized into broader typologies, such as truss or girder bridges. This approach is embodied in the recently introduced concept of Population-Based Structural Health Monitoring (\textit{PBSHM}) (\cite{gosliga2022population,tsialiamanis2023towards,tsialiamanis2024meta,bunce2024population}).

According to structural analysis theory, a population of structures can be represented by graphs that share certain morphological or topological characteristics, despite each instance comprising unique geometric or material properties. Following the logic of graph representations as suggested by \cite{gosliga2021foundations}, Graph Neural Networks (GNNs) can work as efficient representations for \textit{PBSHM}. GNNs are a type of deep learning models that are specifically designed to operate on graph-structured data, comprising a set of nodes and edges that capture interrelations amongst elements of the network. Compared with traditional deep learning models, GNNs present several unique advantages (\cite{sanchez-lengeling2021a}): 1) \textit{Flexibility}: The flexibility of the GNN architecture allows them to handle graphs with varying node counts and diverse connectivity patterns. This characteristic makes them particularly suitable for \textit{PBSHM}-related datasets, where structures within a population share similarities but also possess individual differences. 2) \textit{Accuracy}: GNNs are explicitly designed to handle data that is structured as graphs, which can process data in three different spatial levels: node level, edge level, and graph level. The message passing function of GNNs ensures more effective utilization of spatial information hidden in the data, consequently enhancing accuracy in processing graph-structured data. 3) \textit{Interpretability}: We can naturally correlate data within a given dataset with the node-, edge-, and graph-level features of a GNN. This characteristic facilitates a more straightforward interpretation of both the model itself and the resulting outcomes. 

Due to the above-mentioned advantages of GNNs, a number of studies have already attempted to integrate GNNs with forward analysis tasks. Examples include the use of GNNs to predict the shear stress in wall structures (\cite{dupuy2023modeling,dupuy2024using}), airflow of wind energy systems (\cite{Mylonas2021,duthe2023graph, duthe2023local}), and estimating the main natural frequencies of truss structures given specific environmental conditions (temperature) and different member types (\cite{tsialiamanis2022application}). For SHM, however, inverse problems play a more critical role, where the goal is to identify structural parameters from measurements of structures and, thus, evaluate the condition of a monitored structure (\cite{gallet2022structural}). To date, a paucity of studies involving GNNs for inverse problem solutions is noted. In this study, we take this task on by introducing GNNs for a fundamental SHM task, namely Operation Modal Analysis (OMA), aiming to identify modal properties of a structure based solely on measured vibration response data (\cite{Reynders2012,Brincker2015}). As typically conducted within an OMA setting, the identified modal properties can be used for downstream tasks such as damage identification (\cite{hou2021review, gres2022statistical, gres2023low}) or response prediction (\cite{lai2022neural}).

Given the background introduced above, this study proposes a novel deep learning architecture, with the GNN serving as the pivotal building block, aimed at automatically performing output-only modal identification for structures within a population. Key contributions of this study include:
\begin{itemize}
\item We design the architecture of a GNN-based deep learning model, in which the structural population is modelled by a GNN. This model is trained to capture the meta behavior of the structural population, enabling it to output natural frequencies, damping ratios, and mode shapes from the input that is the Power Spectral Density (PSD) of the vibration acceleration on structure nodes.

\item We adopt the Feature Propagation (FP) algorithm to reconstruct the full-field acceleration PSD using partial measurements of a small subset of structure nodes, in an effort to reflect a realistic SHM context where only sparse dynamic response measurements are available. The proposed GNN-based model can then be used for modal identification based on the reconstructed PSD.

\item We conduct systematic numerical experiments to assess the performance of the proposed GNN-based model. The accuracy and efficiency of the model are thoroughly investigated.
\end{itemize}

\section{Methodology}
\subsection{Problem formulation}
Modal identification of structural systems plays a fundamental role in the context of SHM. In general, the identification task aims at ascertaining the natural frequencies, damping ratios and mode shapes of the monitored structure by processing measurements relating to dynamic response (structural output), such as accelerations, from a finite- and typically sparse- set of Degrees of Freedom (DOFs). OMA schemes rely on use of output-only information under the assumption of broadband and random unmeasured excitation. In keeping true to this requirement, this study employs Gaussian white noise to model the excitation sources of the simulated structures.

To further define the problem, let us denote the available acceleration measurements as $\mathbf{X}(t) \in \mathbb{R}^{N \times P}$, where $t$ represents the time components, $N$ denotes the amount of monitored DOFs, and $P$ denotes the amount of available time samples per signal. In this study, we aim to learn a function $func\left ( \cdot  \right )$ that can identify natural frequencies, damping ratios, and mode shapes of the first $k$ structural modes based on the measured signals, $\mathbf{X}(t)$. Then the identification process can be expressed as:
\begin{equation}
\label{eq1}
\left[ \hat{\mathbf{F}}, \hat{\mathbf{Z}}, \hat{\mathbf{\Phi}} \right] = func(\mathbf{X}(t))
\end{equation}
where $\hat{\mathbf{F}} \in \mathbb{R}_{>0}^{1 \times k}$, $\hat{\mathbf{Z}} \in \mathbb{R}_{>0}^{1 \times k}$, and $\hat{\mathbf{\Phi}} \in \mathbb{R}^{N \times k}$ denote identified natural frequencies (in Hz), damping ratios, and mode shapes, respectively. 

Within an OMA context, $P$ is usually chosen to reflect a large number of time samples, since typically long time series are required for construction of the appropriate spectra. As a consequence, the input of the deep learning model, $\mathbf{X}(t)$, will comprise a high dimensionality, which is often undesirable as it compromises efficiency (\cite{van2009dimensionality}). Therefore, in this study, we reduce the dimension of $\mathbf{X}(t)$ by employing the frequency domain representation of the measured signals, i.e., the Power Spectral Density (PSD). The converted PSD is denoted as ${\mathbf S}(f) \in \mathbb{R}_{\geq 0}^{N \times M}$, where $f$ denotes the frequency components and $M$ is the dimension of the PSD representation. Since the PSD contains only amplitude and not phase information of the time-history signals, only absolute mode shapes $\left | \hat{\mathbf{\Phi}} \right | \in \mathbb{R}_{\geq 0}^{N \times k}$ can be identified from the proposed approach, which are nonetheless still meaningful for SHM tasks. Thus, the problem in this study can be reduced to learn a function $func\left ( \cdot  \right )$ that can be expressed as:
\begin{equation}
\label{eq:problem}
\left[ \hat{\mathbf{F}}, \hat{\mathbf{Z}}, \left|\hat{\mathbf{\Phi}}\right| \right] = func( {\mathbf S}(f))
\end{equation}

\subsection{Model architecture}
\begin{figure}[!htp] 
   \centering 
   \includegraphics[width=13cm]{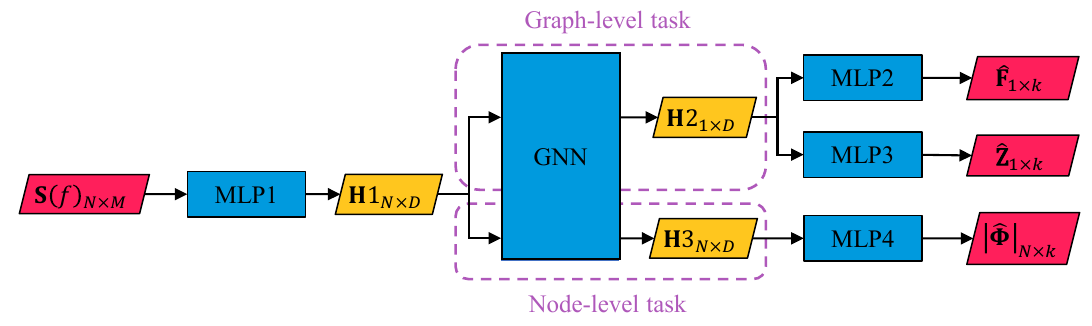}
   \caption{Architecture of the proposed model, in which the model input and output are marked in red color, hidden features are marked in yellow color, and deep learning blocks are marked in blue color.}
   \label{fig:architecture}
\end{figure}

To obtain the modal identification function $func\left ( \cdot  \right )$ shown in Equation \eqref{eq:problem}, we design a deep learning model to learn the mapping between ${\mathbf S}(f)$ and $\left[ \hat{\mathbf{F}}, \hat{\mathbf{Z}}, \left|\hat{\mathbf{\Phi}}\right| \right]$. The architecture of the proposed model is visualized in Figure \ref{fig:architecture}.

As indicated in Figure \ref{fig:architecture}, five deep learning blocks (marked in blue color) are employed in the proposed model. Their description is as follows:
\begin{itemize}

 \item \textit{MLP 1}: The multi-layer perceptron (MLP) 1 block serves as an encoder that compresses the input data $\mathbf S$ ($M$-dimensional) into more compact hidden features $\mathbf H1$ ($D$-dimensional). This encoding process is widely adopted in the design of DL models (\cite{Goodfellow-et-al-2016}). The ablation study (section \ref{sec:ablation}) demonstrates that the usage of this block helps to improve the accuracy of the proposed model.

 \item \textit{GNN}: The main building block in the proposed architecture is the GNN, which is constructed on the basis of engineering intuition. More specifically, as illustrated in Figure \ref{fig:graph}, we can reasonably designate the joint locations of a structure as the nodes of the GNN and the structural elements connecting those nodes (e.g. beams, truss bars) to correspond to the GNN edges. Unlike a common MLP architecture, where the number of neurons is fixed, GNNs can process graphs of various configurations, corresponding to varying numbers of nodes and edges (\cite{zhou2020graph}), which is particularly beneficial for \textit{PBSHM} tasks, since each structure usually comprises a different number of nodes and elements. The flexibility of GNNs comes from their message passing mechanism, which aggregates information from neighboring nodes rather than a fixed set of neurons, with the model parameters being shared across all nodes and edges. Once the skeleton of the GNN is built, the input quantities need to be determined on the basis of available measurements from the monitored systems. For the framework we propose, as shown in Figure \ref{fig:graph}, this involves computation of the PSDs of vibration acceleration signals $\mathbf S$ as well as the absolute mode shapes $\left | \hat{\mathbf{\Phi}} \right |$ as the node-level features of the GNN, while the natural frequencies $\hat{\mathbf{F}}$ and damping ratios $\hat{\mathbf{Z}}$ serve as the graph-level features of the GNN. This study adopts an encoder-decoder architecture, which implies that the GNN does not directly process $\mathbf S$ as the input to produce $\left | \hat{\mathbf{\Phi}} \right |$, $\hat{\mathbf{F}}$, and $\hat{\mathbf{Z}}$ as the output. Instead, the GNN takes the encoded features $\mathbf H1$ as the nodal input and executes two individual tasks, which in Figure \ref{fig:architecture} are correspondingly indicated as the graph-level task and the node-level task. The graph-level task corresponds to the `readout' operation, which is essential for graph-level downstream tasks (\cite{gilmer2017neural}), and serves for generating a single hidden feature $\mathbf H2$ by aggregating node features. In parallel, the node-level task performs the message passing operation, which is the GNN's fundamental mechanism that allows nodes in a graph to exchange information with their neighbors in order to exploit the interrelations that lay latent in the graph-structured data (\cite{wu2020comprehensive}). The product of the node-level task is the hidden feature $\mathbf H3$ on each node.

\begin{figure}[!htp] 
   \centering 
   \includegraphics[width=11cm]{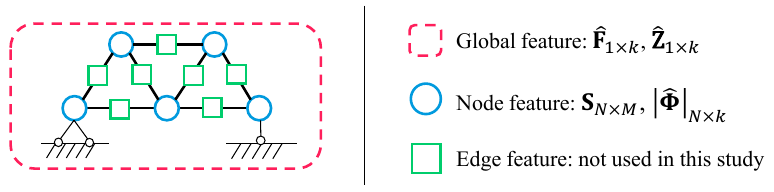}
   \caption{An example of the graph dataset used in this study. A truss structure can be naturally modeled as a graph. See text about the GNN block for more details.}
   \label{fig:graph}
\end{figure}

 \item \textit{MLP 2\&3\&4}: Finally, three MLPs are employed as the decoder architecture in order to project the $D$-dimensional hidden features back to the physical domain, yielding the modal properties of the first $k$ modes. The MLP 2 and MLP 3 blocks decode the graph-level hidden feature $\mathbf H2$, and thus output the graph-level modal properties, which are natural frequencies $\hat{\mathbf{F}}$ and damping ratios $\hat{\mathbf{Z}}$, respectively. The MLP 4 block decodes the node-level hidden feature $\mathbf H3$ and outputs the mode shapes $\left | \hat{\mathbf{\Phi}} \right |$ of every node.
 
\end{itemize}

\subsection{Loss function}

To train our model, we employ an objective function based on the Mean Squared Error (MSE):

\begin{equation}
\label{eq:loss}
     \mathcal{L}  =  \frac{\lambda_1}{kN}\sum_{i=1}^{N}\sum_{j=1}^{k}\left ( \left | \mathbf{\Phi} \right |_{i,j} - \hat{\left | \mathbf{\Phi} \right |}_{i,j} \right)^{2} + \frac{\lambda_2}{k}\sum_{j=1}^{k}\left ( \frac{\hat{\mathbf{F}}_{j}}{\mathbf{F}_{j}}-1\right )^{2} + \frac{\lambda_3}{k}\sum_{j=1}^{k}\left ( \frac{\hat{\mathbf{Z}}_{j}}{\mathbf{Z}_{j}} - 1\right )^{2}
 \end{equation}
where ${\mathbf{F}}$, ${\mathbf{Z}}$, and $\left | {\mathbf{\Phi}} \right |$ are the target natural frequencies, damping ratios, and absolute mode shapes, respectively. These quantities are known during training, and the methods for deriving them will be elaborated in Section \ref{sec:framework}. $\lambda_1$, $\lambda_2$, and $\lambda_3$ are coefficients that are adopted to balance the contribution of the individual components related to ${\mathbf{F}}$, ${\mathbf{Z}}$, and $\left | {\mathbf{\Phi}} \right |$ to the loss function. In this study, we determine $\lambda_1$, $\lambda_2$, and $\lambda_3$ by trial and error. 

It is also noteworthy that, in this study the PSDs (model input) and absolute mode shapes (model output) are normalized by scaling their amplitudes to a maximum of 1 unit before training to boost the generalization ability of the deep learning model. However, natural frequencies and damping ratios cannot be normalized by max normalization because their original values matter, and this can cause imbalance during model training. For instance, natural frequencies are generally much larger than damping ratios, so the loss term regarding natural frequencies will outweigh the damping ratio loss if they are not normalized. To diminish this possible imbalance, as shown in the loss function, we normalize natural frequencies and damping ratios by dividing the estimated values (model output, which are ${\mathbf{\hat{F}}}$, ${\mathbf{\hat{Z}}}$) by their corresponding target values (${\mathbf{F}}$, ${\mathbf{Z}}$).


\subsection{Framework}
\label{sec:framework}
\begin{figure}[!htp] 
   \centering 
   \includegraphics[width=14.3cm]{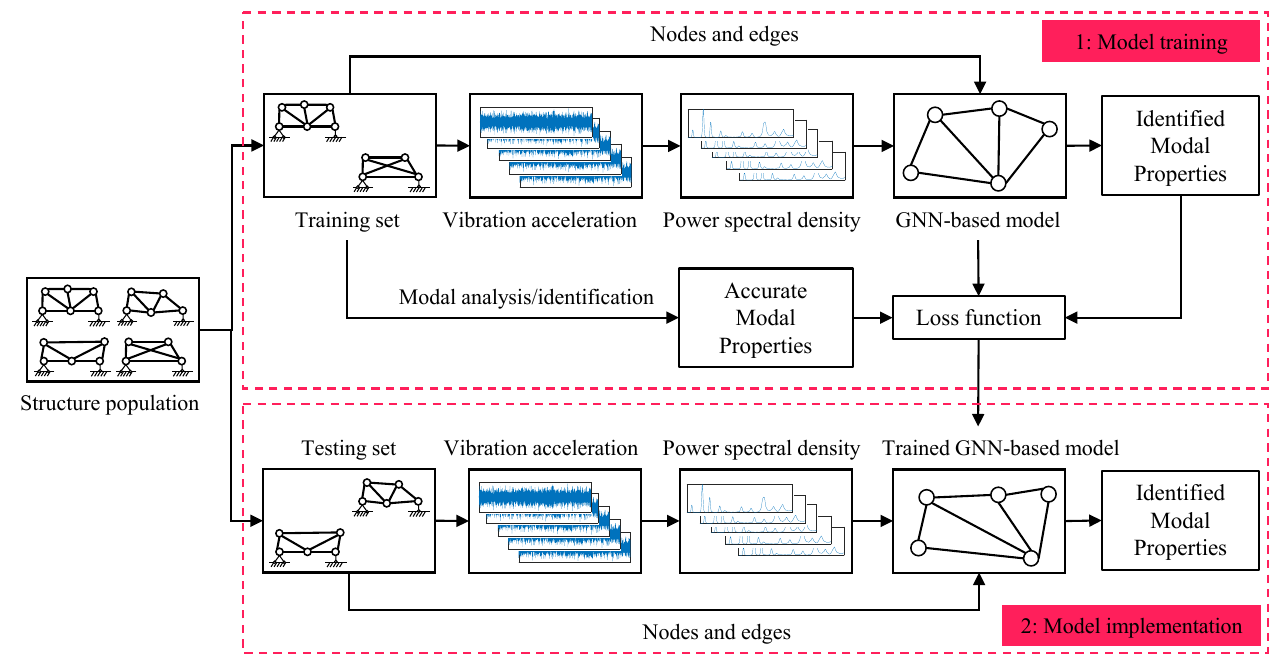}
   \caption{Framework of using the GNN-based model for population-based structural modal identification}
   \label{fig:framework}
\end{figure}

On the basis of above-mentioned problem formulation, model architecture, and loss function, the proposed GNN-based \textit{PBSHM} framework for modal identification is summarized in Figure \ref{fig:framework}.
As shown in Figure \ref{fig:framework}, the framework is divided into two stages, namely model training and model implementation. In the model training stage, a subset of the employed structural population is used as the training set. To establish such a structural population, numerical simulation is necessary because for real-world datasets it is impossible to monitor the dynamic response of all DOFs of structures, while the full-field measurements and mode shapes are required during the training stage. For the simulated population, structures within it are randomly generated and modelled by the finite element method. Then the modal properties of each structure are calculated by means of eigenvalue analysis. To make the simulated population more realistic, some of the members could be modelled using the information from available monitored structures in the real world, and model updating could be performed based on the measurements to enhance the fidelity of the simulation. Nonetheless, in this study we only verify the proposed model with pure numerical experiments, so the model updating part is not involved. More details about dataset generation can be seen in Section \ref{sec:dataset} below.

After the training is completed, the trained model can be readily used to automatically identify the modal properties of other unseen structures within the structural population, even though their configuration may be different to what is met in the training set, i.e., correspondent to a different number of nodes and edges and connectivity. This identification process is end-to-end, automatic, and fast, delivering a main advantage over existing modal identification schemes. A more detailed comparison can be found in Section \ref{sec:comparison}.

\section{Numerical experiments}
In this section, we present the conducted numerical experiments, which were designed to answer the following research questions:
\begin{itemize}
\item Question 1: Which type of GNN is more suitable for the population-based OMA task?
\item Question 2: How do the deep learning modules in the model architecture, especially the GNN module, contribute to the overall model performance?
\item Question 3: Can the proposed model effectively handle real-world challenges such as incomplete measurements, measurement noise, and limited number of structures in the dataset?
\item Question 4: What are the advantages and disadvantages of the proposed model versus existing modal identification methods?
\item Question 5: Does the proposed model demonstrate sufficient generalization ability across diverse structural populations?
\end{itemize}

\subsection{Dataset description}
\label{sec:dataset}
\begin{figure}[!htp] 
   \centering 
   \includegraphics[width=11.5cm]{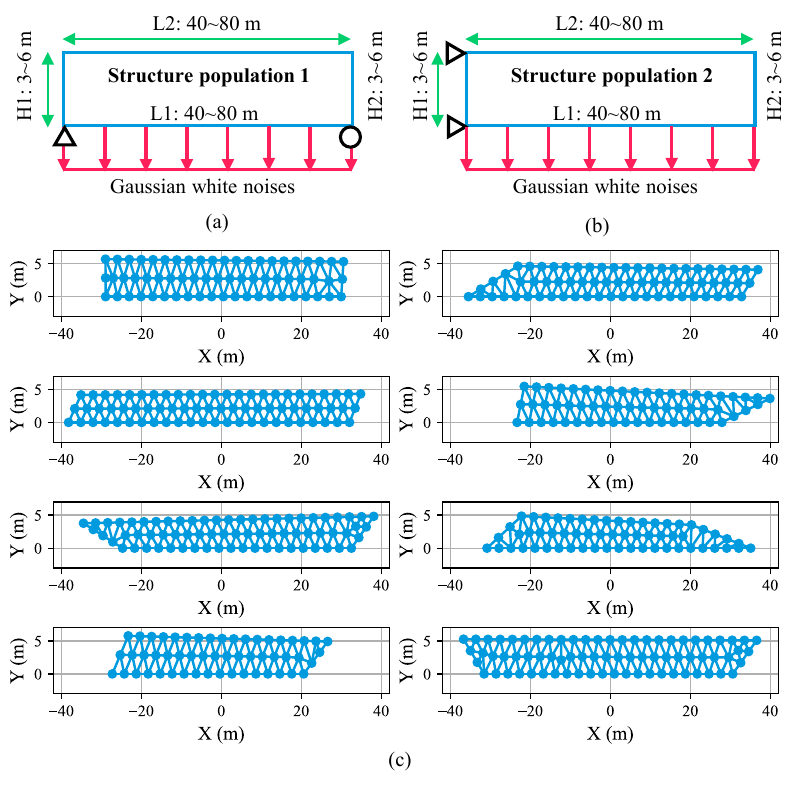}
   \caption{Visualization of the two generated datasets: (a) Geometric configuration meant to approximate a simply-supported truss population; (b) Geometric configuration serving to approximate a cantilevered truss population; (c) Some representative truss samples from the two datasets (only nodes and elements are displayed)}
   \label{fig:dataset}
\end{figure}

The extraction of datasets from real-world structural populations is non-trivial, although recent efforts in Asia \citep{Fernando2018Technical} and Europe \citep{limongelli2024bridge} do involve denser instrumentation of structural populations (such as bridges). These endeavors are expected to deliver datasets that can support \textit{PBSHM} tasks.
However, these datasets have not been made public yet, so in this work, we initiate from simulated scenarios for two datasets that correspond to two different truss populations. The first truss population consists of 500 trusses assumed to be arranged within a trapezoidal boundary that are meant to approximate the geometrical configuration of simply-supported beam structures. Each truss is generated by randomly meshing the trapezoidal area with Delaunay triangles (\cite{persson2004simple}). The first 400 trusses are used to train the model, and the remaining 100 trusses are used to test the model. The geometric boundary and some generated truss examples are shown in Figure~\ref{fig:dataset}(a) and Figure~\ref{fig:dataset}(c), respectively. Based on the generated geometric configurations, corresponding finite element models are straightforwardly created using truss elements. The density and area of truss elements are constantly set as 8015 kg/m$^3$ and 0.5 m$^2$, respectively. In order to reflect varying material properties, the Young's modulus of the employed truss elements is set as a random number ranging from 100 GPa to 300 GPa. Figure~\ref{fig:dataset}~(a) further indicates the boundary conditions (simply-supported type) and the external excitation (Gaussian white noise) that are imposed on the bottom boundary. Linear time history analyses (Newmark-$\beta$ method) are then performed to obtain the in-plane vertical nodal acceleration (response) time series, of 60 seconds duration, sampled at a time step of 0.005 second (200 Hz). Welch's method is next applied to convert the time series into PSDs, which serves as the input of the suggested DL architecture. An eigenvalue analysis is further conducted on all 500 finite element models, generating the reference modal properties of the primary (first 4 in our study) modes for model training and testing. It should be noted that, for model training, the complete full-field nodal acceleration measurements are used, whereas the trained model is tested with both the complete and incomplete measurements. More details in this regard are offered in Section \ref{sec:incomplete}.

In order to further test the generalization ability of the proposed model, we simulate a separate structural population comprised of 100 samples that are meant to approximate the geometry of cantilevered trusses. This population is unseen for the learning \textit{PBSHM} model, which is trained on the simply-supported structural population. As illustrated in Figure~\ref{fig:dataset}(b), the second population only differs from the first population in the structural supports. Except for the structural supports, all the other settings, including geometric boundaries, material properties, and excitation, are the as same as those of the first structural population. More details in this regard can be found in the following Section \ref{sec:population}.

\subsection{Implementation details}
The environment we implement our model is set up in PyTorch 2.0.1, CUDA 11.7, and DGL 1.11. The Adam optimizer is adopted to train our model with the default setting (learning rate: 0.001, first momentum decaying parameter: 0.9, second momentum decaying parameter: 0.999), and we use a NVIDIA GeForce RTX 3060 to accelerate the batch training process. The training batch size is 64, with 5,000 epochs deployed. The MLPs and GNNs used in our study all have three layers, and the adopted dimension is 64 for every layer. In the loss function, we set the coefficients $\lambda_1$, $\lambda_2$, and $\lambda_3$ as 2, 1, and 1, respectively. More implementation details can be found in our GitHub repository (\cite{Jian2024}), which will be made public upon publication of this work.

\subsection{Effects of different GNN models}
Depending on the type of message passing function that is adopted, different GNN modeling instances can be created. In this study, we compare three commonly-used GNN models, namely Graph Convolutional Networks (GCNs, proposed by \cite{kipf2016semi}), Graph Attention Networks (GATs, proposed by \cite{velivckovic2017graph}), and GraphSAGE (proposed by \cite{hamilton2017inductive}). To control the comparison, we only replace the GNN block in the model architecture shown in Figure \ref{fig:architecture} with different GNN models. Then different models are trained using the K-fold cross validation, which means we evenly divide the training set that consists of 400 trusses into 5 folds. In each training process, 4 folds of trusses (320 in number) are used to train the model, and the remaining 1 fold of trusses (80 in number) are used to validate the model. We train the model for 5 times, so that all the 5 folds can serve as the validation set in turn. The mean values and Standard Deviation (SD) of the final validation loss and training time of the 5 training processes are shown in Table \ref{tab:GNNmodel}.
\begin{table}[!htp] 
\centering
\caption{Comparison between the different GNN models}
\label{tab:GNNmodel}
\begin{tabular}{ccccccc}
\hline
       & \multicolumn{2}{c}{GCN} & \multicolumn{2}{c}{GAT} & \multicolumn{2}{c}{GraphSAGE} \\
        & Mean   & SD    & Mean   & SD    & Mean     & SD        \\ \hline
Final validation loss & 0.076  & 0.019 & 0.066  & 0.012  & \textbf{0.059}   & \textbf{0.010}   \\
Training time (s)  & 519.696 & \textbf{10.293}  & 699.911 & 14.639  & \textbf{443.838}  & 11.881   \\ \hline
\multicolumn{7}{l}{\small {*Indicators showing better performance are shown in bold.}} \\
\end{tabular}
\end{table}

Based on Table~\ref{tab:GNNmodel}, we observe that the GraphSAGE model demonstrates superior performance, since both its final validation loss and required training time are statistically smaller than the GCN and GAT model. This is because, according to \cite{hamilton2017inductive}, the GraphSAGE exhibits inductive learning capabilities, scalability via neighborhood sampling, and flexibility in aggregation methods. These features render it particularly suitable for large-scale and dynamic graphs such as those employed herein within the \textit{PBSHM} context, where efficiency and the ability to handle new nodes without re-training are crucial. GCNs are less scalable and inductive, while GATs offer sophisticated attention mechanisms at the cost of increased computational complexity. As a result, hereafter we choose to adopt the GraphSAGE model as the GNN block, unless mentioned otherwise.

\begin{figure}[!htp] 
   \centering 
   \begin{tabular}{ccc}
   \includegraphics[width=4.5cm]{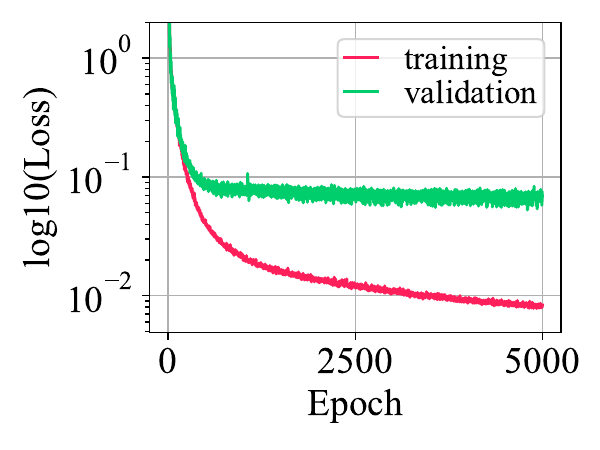} & \includegraphics[width=4.5cm]{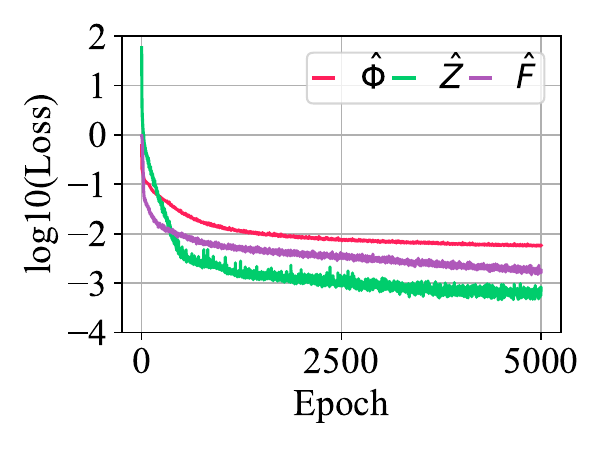} & \includegraphics[width=4.5cm]{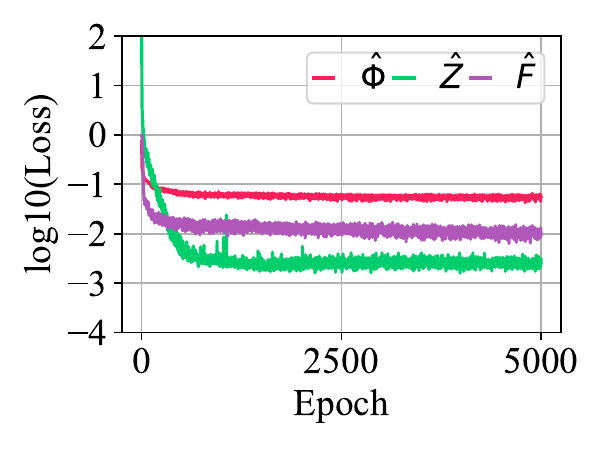} \\
   (a) & (b) & (c) \\
   \end{tabular}
    \caption{Loss curves of one training process among the K-Fold Cross-Validation: (a) Total training and validation loss; (b) Different loss terms in the training loss; (c) Different loss terms in the validation loss}
   \label{fig:loss}
\end{figure}

Now that the particular type of GNN models is determined, it is necessary to elaborate on the details of the training and testing process in order to comprehensively report on the proposed model. Figure \ref{fig:loss} shows the loss curves corresponding to one training process among the K-Fold Cross-Validation. The total loss (Equation \ref{eq:loss}), the loss terms regarding mode shapes ($\mathbf{\hat{\Phi}}$), damping ratios ($\mathbf{\hat{Z}}$), and natural frequencies ($\mathbf{\hat{F}}$) are all displayed.


Figure \ref{fig:loss} allows for the following observations:
\begin{itemize}

\item The training loss results smaller than the validation loss in Figure \ref{fig:loss}(a), which is reasonable as the validation loss is not used in backpropagation for updating the model weights.

\item Nonetheless, the validation loss is introduced for deciding when to terminate the training process and for facilitating the comparison of different GNN models in this subsection as well as different model architecture in the ablation study below. Although the training loss appears to continually decrease in Figure \ref{fig:loss}(b), the validation loss appears to have converged in Figure \ref{fig:loss}(c), which promoted termination of the training process after 5,000 epochs.

\end{itemize}

\begin{figure}[!htp] 
   \centering 
   \begin{tabular}{ccc}
   \includegraphics[width=4.52cm]{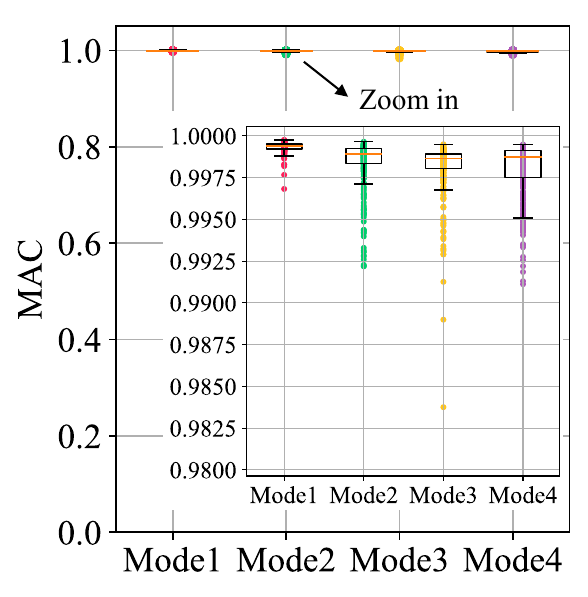} & \includegraphics[width=4.5cm]{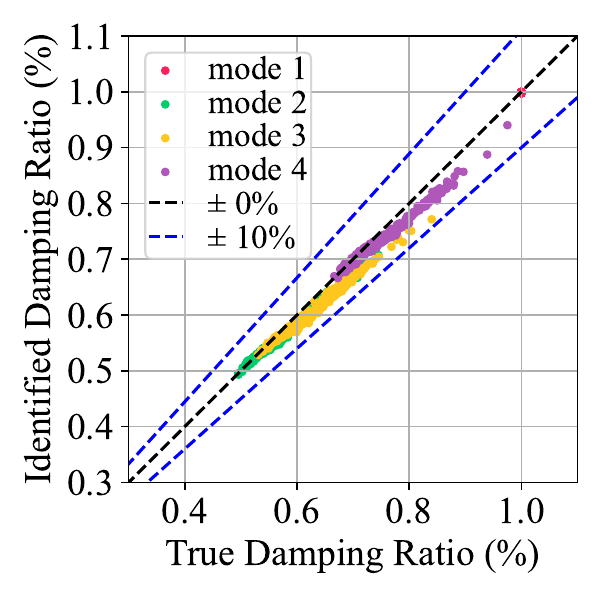} & \includegraphics[width=4.5cm]{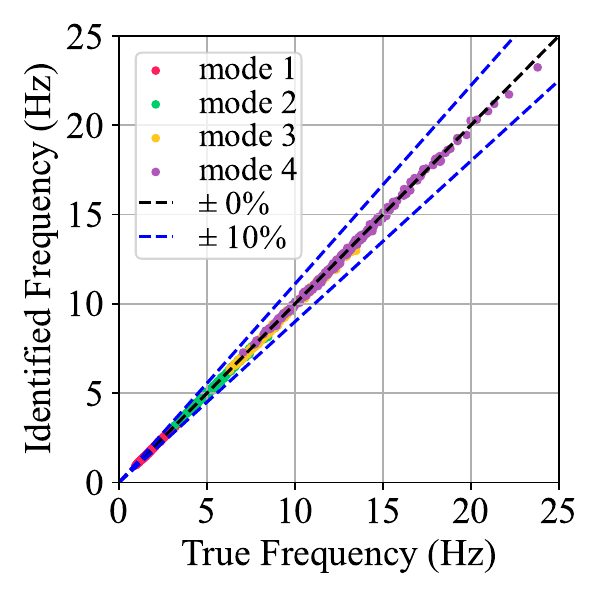} \\
   (a) & (b) & (c) \\
   \end{tabular}
    \caption{Performance of the trained model on the training set: (a) Box plot of Modal Assurance Criterion (MAC) values of identified mode shapes; (b) Scatter plot of identified damping ratios; (c) Scatter plot of identified natural frequencies}
   \label{fig:train_popu1}
\end{figure}

\begin{figure}[!htp] 
   \centering 
   \begin{tabular}{ccc}
   \includegraphics[width=4.5cm]{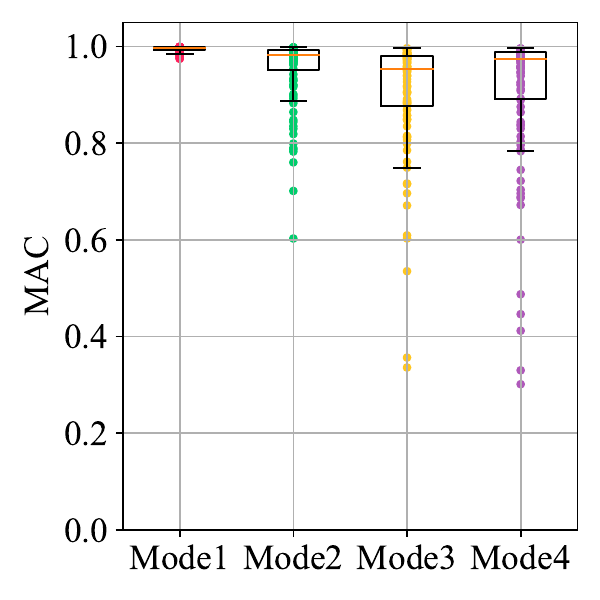} & \includegraphics[width=4.5cm]{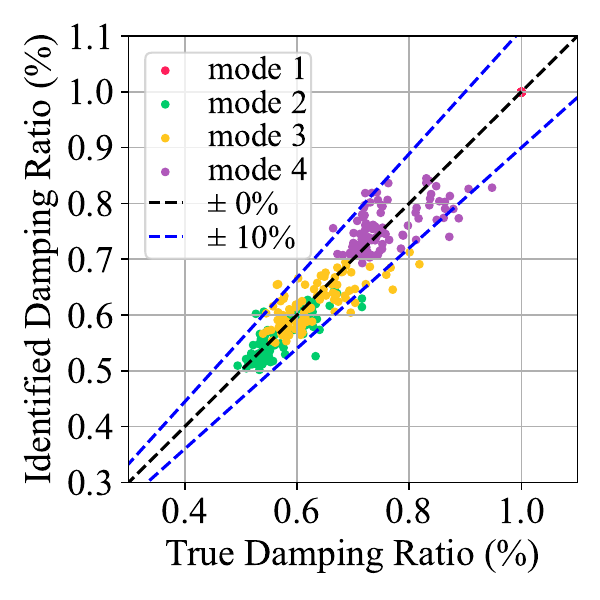} & \includegraphics[width=4.5cm]{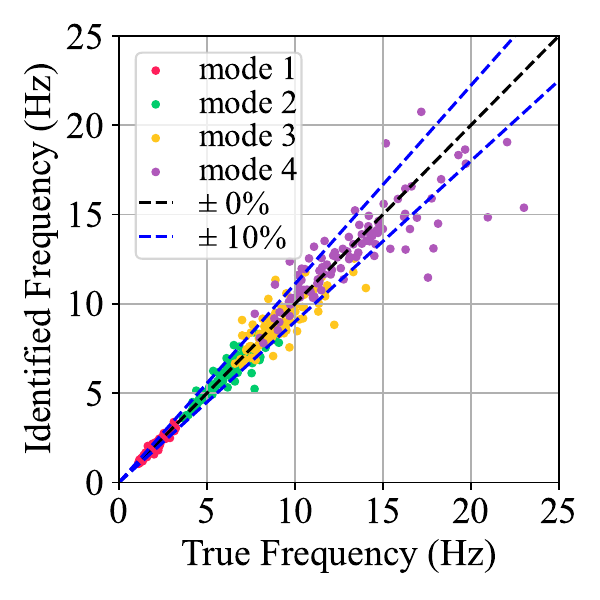} \\
   (a) & (b) & (c) \\
   \end{tabular}
    \caption{Performance of the trained model on the testing set: (a) Box plot of MAC values of identified mode shapes; (b) Scatter plot of identified damping ratios; (c) Scatter plot of identified natural frequencies}
   \label{fig:test_popu1}
\end{figure}

After the training is completed, we firstly use the trained model to identify the modal properties of truss configurations lying within the simply-supported structural population. Identification results of training set (320 trusses) and testing set (100 trusses) are visualized in Figure \ref{fig:train_popu1} and Figure \ref{fig:test_popu1}, respectively, in which each dot represents the identification result per individual truss per mode and different order of modes are distinguished by different colors. Figure \ref{fig:train_popu1}(a) and Figure \ref{fig:test_popu1}(a) show the Modal Assurance Criterion (MAC) values of identified mode shapes, of which is defined by Equation \eqref{eq:MAC}, along with their box plots. The closer the MAC value is to 1, the higher the accuracy of the mode shape identification.

\begin{equation}
    \label{eq:MAC}
    \text{MAC}(|\hat{\phi}|, |\phi|) = \frac{ \left ( |\hat{\phi}|^T |\phi| \right )^2}{(|\hat{\phi}|^T |\hat{\phi}|)(|\phi|^T |\phi|)}
\end{equation}
where $|\hat{\phi}|$ and $|\phi|$ are vectors of the identified and target absolute mode shapes, respectively.

The identified damping ratios and natural frequencies are shown in Figure \ref{fig:train_popu1}(b), Figure \ref{fig:test_popu1}(b), Figure \ref{fig:train_popu1}(c), and Figure \ref{fig:test_popu1}(c), respectively. In these figures, the further a point lies from the $\pm0\%$-error baseline, the larger the weighing error. Two extra reference lines representing the $\pm10\%$ relative error are also plotted in the figure to further evaluate the identification accuracy and reliability.

Figure \ref{fig:train_popu1} and Figure \ref{fig:test_popu1} allow for the following findings:
\begin{itemize}

\item The training process is deemed as successful, since the majority of evaluation points, as reported in Figure \ref{fig:train_popu1}(a), lie close to 1, and the points plotted in Figure \ref{fig:train_popu1}(b) and Figure \ref{fig:train_popu1}(c) are scattered near the $\pm0\%$-error baseline.

\item The identification results on the testing set are noticeably worse than those on the training set. Despite a few instances of very low identification accuracy, the errors generally lie below 10\%, which is considered to be acceptable, particularly considering that the identification process is fast and completely end-to-end. To quantitatively understand the performance of the model on the testing set, Table \ref{tab:PI_popu1} presents the statistics of the MAC values of identified mode shapes and the relative errors (\%) for the identified damping ratios and natural frequencies. As shown, the average MAC values for the identified mode shapes for all four modes are above 0.9, and the average errors corresponding to the damping ratios and natural frequencies are merely 1\% - 2\%. This indicates that the trained model statistically performs well in the modal identification of unseen structures among the same structural population. 
\end{itemize}

\begin{table}[!htp] 
\centering
\caption{Performance indicators on the testing set from the simply-supported structural population}
\label{tab:PI_popu1}
\begin{tabular}{cccccccccc}
\hline
& \multicolumn{3}{c}{MAC of $\hat{\mathbf{\Phi}}$} & \multicolumn{3}{c}{Errors (\%) of $\hat{\mathbf{Z}}$} & \multicolumn{3}{c}{Errors (\%) of $\hat{\mathbf{F}}$} \\
& Mean   & SD  & Min   & Mean   & SD  & Max  & Mean     & SD   & Max      \\ 
\hline
Mode 1 & 0.995                & 0.006                & 0.975                & -0.069               & 0.033                & 0.132                & -0.445               & 6.740                & 24.915               \\
Mode 2 & 0.952                & 0.072                & 0.603                & -1.853               & 4.479                & 16.905               & -1.351               & 7.201                & 32.007               \\
Mode 3 & 0.902                & 0.124                & 0.336                & -0.608               & 5.974                & 16.259               & -1.804               & 9.180                & 30.053               \\
Mode 4 & 0.905                & 0.148                & 0.302                & 0.304                & 5.867                & 15.076               & -0.765               & 10.911               & 34.681               \\
\hline
\multicolumn{10}{l}{\small {*$\hat{\mathbf{\Phi}}$/$\hat{\mathbf{Z}}$/$\hat{\mathbf{F}}$: identified mode shapes/damping ratios/natural frequencies; SD: Standard Deviation.}} \\
\end{tabular}
\end{table}

\subsection{Ablation study}
\label{sec:ablation}
To justify the model architecture shown in Figure~\ref{fig:architecture}, we perform an ablation study by comparing the following model variants: 
\begin{itemize}
\item \textbf{No encoder}: remove the encoder (MLP1) and retain the remaining blocks in the model architecture. 
\item \textbf{No GNN for mode shape identification}: replace the GNN block with a MLP when converting the hidden features $\mathbf{H1}_{N \times D}$ to $\mathbf{H3}_{N \times D}$ in the model architecture. It should be noted that the GNN cannot be completely removed from the architecture because the identification of damping ratios and natural frequencies require the graph readout operation. Without the GNN, the MLP that has fixed input dimensions will not be able to convert the hidden features $\mathbf{H1}_{N \times D}$ to $\mathbf{H2}_{1 \times D}$, making the architecture inapplicable to structures that posse different number of nodes among a structural population.
\end{itemize}
To achieve this comparison, we train the originally proposed model and the two variants for 5 times with the K-Fold Cross-Validation method. The mean values and standard deviation (SD) of the final validation loss and training time of the 5 training processes are shown in Table \ref{tab:ablation}.
\begin{table}[!htp] 
\centering
\caption{Comparison for ablation study}
\label{tab:ablation}
\begin{tabular}{ccccccc}
\hline
& \multicolumn{2}{c}{No encoder} & \multicolumn{2}{c}{No GNN} & \multicolumn{2}{c}{Original model} \\
        & Mean   & SD    & Mean   & SD    & Mean     & SD        \\ \hline
Final validation loss & 0.075  & 0.024 & 0.085  & 0.021  & \textbf{0.059}   & \textbf{0.010}   \\
Training time (s)  & 396.511  & \textbf{5.434}  & \textbf{287.386} & 9.391  & 443.838  & 11.881  \\ \hline
\multicolumn{7}{l}{\small {*Indicators showing better performance are shown in bold.}} \\
\end{tabular}
\end{table}

Based on Table \ref{tab:ablation}, we can draw the following conclusions:
\begin{itemize}
\item The final validation loss indicates that the original model, whose architecture is illustrated in Figure \ref{fig:architecture}, is the optimal choice as this exhibits the lowest validation loss, which justifies the design of the proposed model. The validation loss for the case of 'No GNN' model is not only higher than the other two models in this ablation study, but also higher than GCN and GAT models, confirming that the GNN does have an advantage of processing graph-structured data over the MLP.
\item Removing the encoder and the GNN from the architecture can reduce the complexity of the model and thus shorten the training time, but this comes at the expense of accuracy. In the context of the modal identification problem, the identification accuracy far outweighs the training time. Hence, the original model is preferable.
\end{itemize}

\subsection{Effects of incomplete measurements}
\label{sec:incomplete}
In practice, the DOFs of civil engineering structures usually significantly outnumber the measurement points, so it is essential to investigate whether the proposed model can identify complete mode shapes from incomplete measurements. As stated previously, we construct the GNN corresponding to the analytical model of a structure. For those structural nodes that have no measurements, their node features are unknown, and the GNN model used in this study cannot work with unknown node features. To make the proposed model work in the case of unknown nodes features, we introduce the Feature Propagation (FP) algorithm (\cite{rossi2022unreasonable}), to fill the unknown features, which are nodal acceleration PSD, on the basis of known features and the graph structure. Applying FP transforms the incomplete measurements into complete ones, enabling our GNN-based model to conduct modal identification.

\begin{figure}[!htp] 
   \centering 
   \begin{tabular}{c}
   \includegraphics[width=14.5cm]{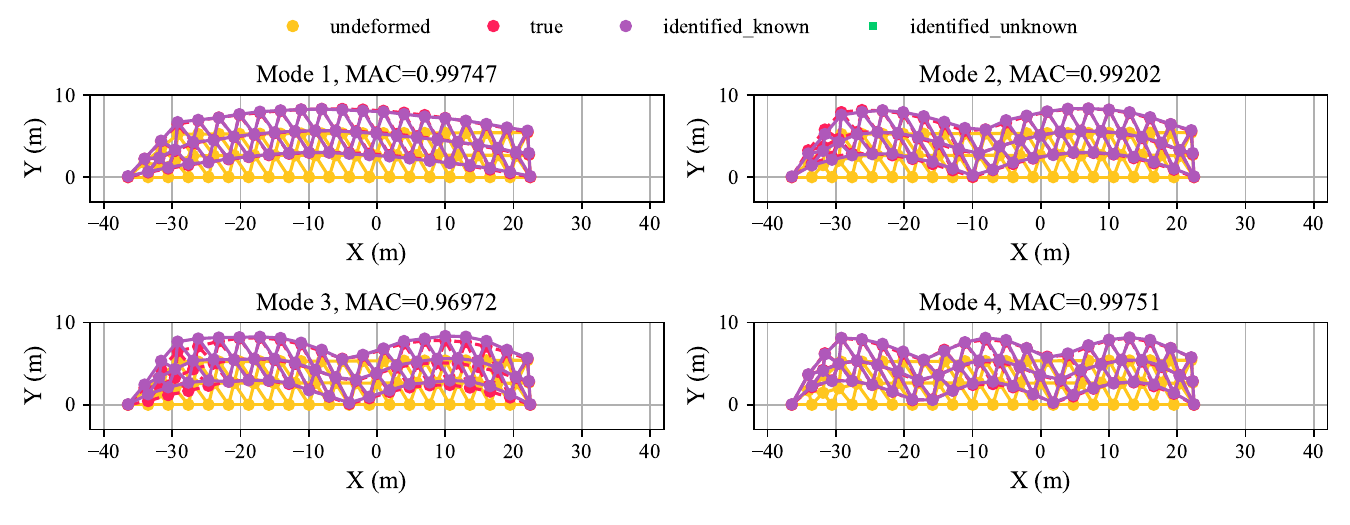} \\
   (a)  \\
   \includegraphics[width=14.5cm]{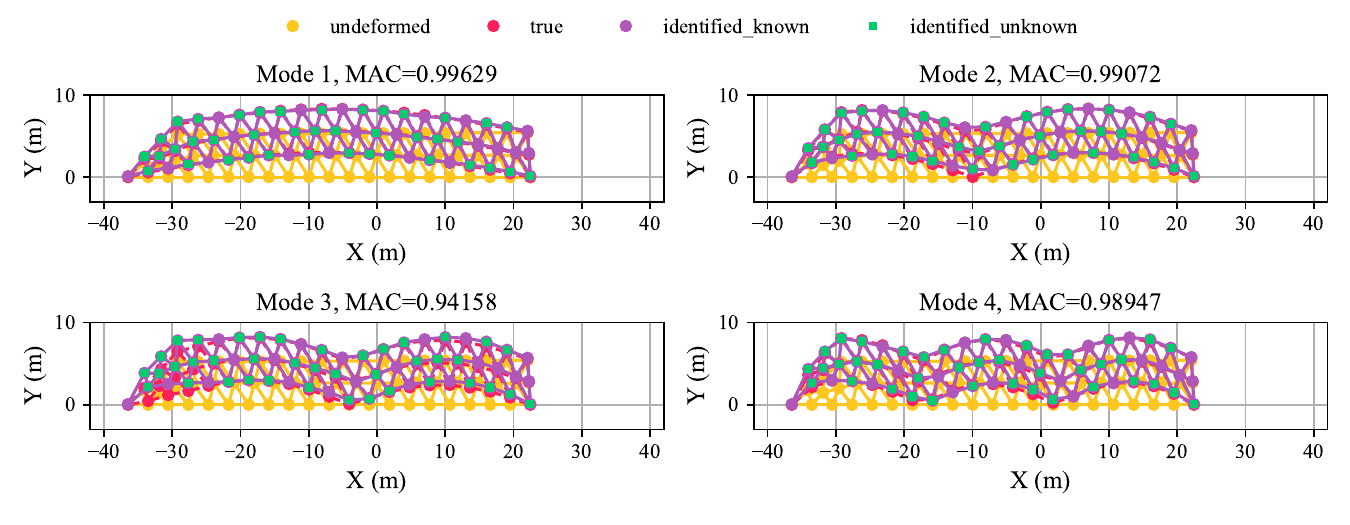} \\
   (b)  \\
   \includegraphics[width=14.5cm]{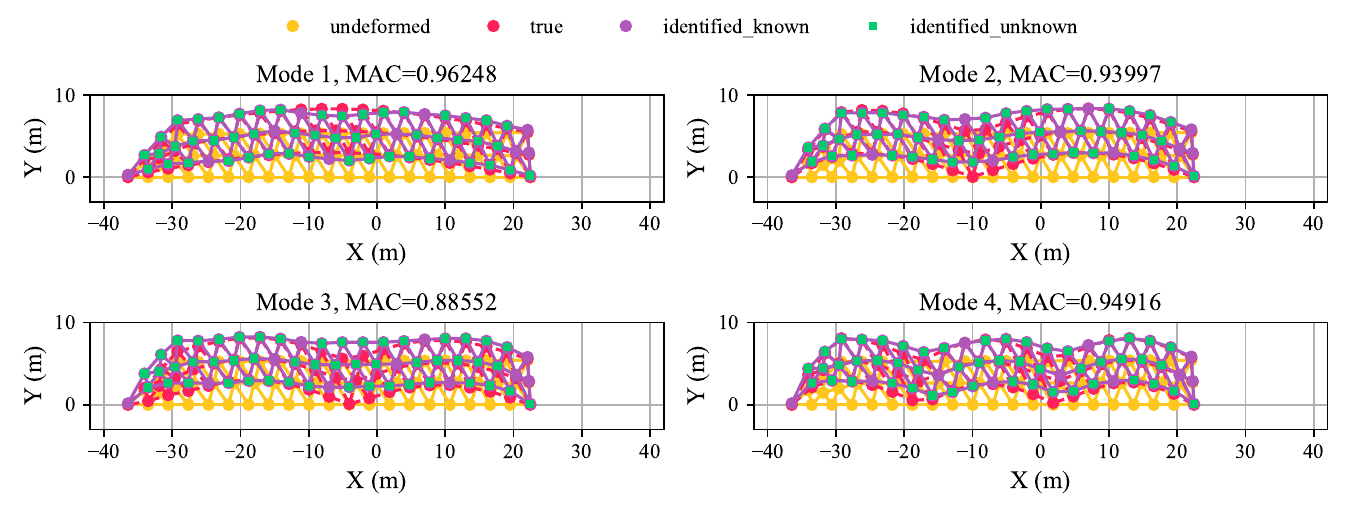} \\
   (c)  \\
   \end{tabular}
    \caption{Mode shape identification results of one truss example with incomplete measurements: (a) 0\% node features are unknown; (b) 66\% nodes node features are unknown; (c) 82\% node features are unknown}
   \label{fig:incomplete}
\end{figure}

Figure \ref{fig:incomplete} illustrates the mode shape identification results of a truss example in the testing set with different assumed ratios of unknown node features, meaning under availability of a lower amount of sensors. We can see that even when as much as 82\% of node features are unknown, which corresponds to a realistic instrumentation setting, the proposed model can still identify mode shapes with an acceptable accuracy.

For further evaluation, Table \ref{tab:PI_popu1_missing66} and Table \ref{tab:PI_popu1_missing82} offer statistics of the MAC values of identified mode shapes and the relative errors (\%) of identified damping ratios and natural frequencies, under  66\% and 82\% unknown node features, respectively.
\begin{table}[!htp] 
\centering
\caption{Performance indicators on the testing set from the simply-supported structural population - 66\% node features are unknown}
\label{tab:PI_popu1_missing66}
\begin{tabular}{cccccccccc}
\hline
& \multicolumn{3}{c}{MAC of $\hat{\mathbf{\Phi}}$} & \multicolumn{3}{c}{Errors (\%) of $\hat{\mathbf{Z}}$} & \multicolumn{3}{c}{Errors (\%) of $\hat{\mathbf{F}}$} \\
& Mean   & SD  & Min   & Mean   & SD  & Max  & Mean     & SD   & Max      \\ 
\hline
Mode 1 & 0.988                & 0.010                & 0.940                & -0.063               & 0.038                & 0.143                & -2.125               & 6.965                & 25.400               \\
Mode 2 & 0.920                & 0.097                & 0.589                & -1.465               & 4.838                & 17.971               & -2.305               & 7.551                & 33.343               \\
Mode 3 & 0.874                & 0.143                & 0.046                & 0.011                & 6.346                & 18.002               & -2.524               & 9.078                & 29.748               \\
Mode 4 & 0.860                & 0.161                & 0.245                & 1.488                & 6.283                & 17.766               & -1.073               & 11.150               & 35.750               \\

\hline
\multicolumn{10}{l}{\small {*$\hat{\mathbf{\Phi}}$/$\hat{\mathbf{Z}}$/$\hat{\mathbf{F}}$: identified mode shapes/damping ratios/natural frequencies; SD: Standard Deviation.}} \\
\end{tabular}
\end{table}

\begin{table}[!htp] 
\centering
\caption{Performance indicators on the testing set from the simply-supported structural population - 82\% node features are unknown}
\label{tab:PI_popu1_missing82}
\begin{tabular}{cccccccccc}
\hline
& \multicolumn{3}{c}{MAC of $\hat{\mathbf{\Phi}}$} & \multicolumn{3}{c}{Errors (\%) of $\hat{\mathbf{Z}}$} & \multicolumn{3}{c}{Errors (\%) of $\hat{\mathbf{F}}$} \\
& Mean   & SD  & Min   & Mean   & SD  & Max  & Mean     & SD   & Max      \\ 
\hline
Mode 1 & 0.971                & 0.018                & 0.906                & -0.064               & 0.040                & 0.150                & -2.945               & 7.885                & 36.507               \\
Mode 2 & 0.847                & 0.122                & 0.422                & -1.520               & 5.435                & 19.572               & -2.196               & 7.370                & 30.076               \\
Mode 3 & 0.793                & 0.160                & 0.070                & 0.099                & 7.337                & 22.738               & -2.357               & 8.905                & 27.269               \\
Mode 4 & 0.744                & 0.192                & 0.136                & 1.757                & 6.942                & 19.019               & -0.566               & 12.228               & 35.295               \\
\hline
\multicolumn{10}{l}{\small {*$\hat{\mathbf{\Phi}}$/$\hat{\mathbf{Z}}$/$\hat{\mathbf{F}}$: identified mode shapes/damping ratios/natural frequencies; SD: Standard Deviation.}} \\
\end{tabular}
\end{table}

By comparing Table \ref{tab:PI_popu1_missing66}, Table \ref{tab:PI_popu1_missing82}, and Table \ref{tab:PI_popu1} (no unknown node features), we can conclude that:
\begin{itemize}
\item As the ratio of unknown node features increases, the identification accuracy drops, which is naturally expected. Moreover, the accuracy of identifying high-order mode shapes drops more than low-order mode shapes due to the greater complexity of high-order mode shapes. 
\item Compared with the identification of mode shapes, the identification of natural frequencies and damping ratios are much less influenced by the increase of unknown node measurements. This is reasonable because, in theory, natural frequencies and damping ratios can be successfully identified with only a few node measurements, whereas accurate identification of mode shapes requires measurements on more structural nodes. This observation implies that the trained model successfully learns the mechanism of modal identification.
\item The modal identification accuracy is statistically acceptable despite being as high as 82\% unknown node features, which manifests the robustness of the proposed model.
\end{itemize}

\subsection{Effects of measurement noise}
In real-world applications, noise corruption is inevitable when acquiring signals that measure the vibration acceleration of structures. To test our model's sensitivity to measurement noise, we artificially pollute the acceleration time series generated by the testing set by injecting white Gaussian noise of which the Root Mean Square (RMS) is 10\% of the signal power. Then we use the Welch's method to obtain the PSD of the polluted time series and input these to the model trained with noise-free data for modal identification. Table \ref{tab:PI_popu1_noisy} reports the statistics of the MAC values of identified mode shapes and the relative errors (\%) of identified damping ratios and natural frequencies on the noise-polluted testing set.
\begin{table}[!htp] 
\centering
\caption{Performance indicators on the testing set from the simply-supported structural population - adding 10\% white Gaussian noise}
\label{tab:PI_popu1_noisy}
\begin{tabular}{cccccccccc}
\hline
& \multicolumn{3}{c}{MAC of $\hat{\mathbf{\Phi}}$} & \multicolumn{3}{c}{Errors (\%) of $\hat{\mathbf{Z}}$} & \multicolumn{3}{c}{Errors (\%) of $\hat{\mathbf{F}}$} \\
& Mean   & SD  & Min   & Mean   & SD  & Max  & Mean     & SD   & Max      \\ 
\hline
Mode 1 & 0.994                & 0.005                & 0.973                & -0.066               & 0.040                & 0.141                & -11.717              & 8.775                & 37.612               \\
Mode 2 & 0.944                & 0.076                & 0.610                & -2.657               & 4.607                & 17.549               & -8.856               & 7.669                & 38.872               \\
Mode 3 & 0.872                & 0.131                & 0.278                & -0.249               & 6.113                & 15.956               & -8.752               & 9.126                & 37.960               \\
Mode 4 & 0.876                & 0.149                & 0.330                & 1.515                & 6.260                & 15.042               & -7.107               & 11.747               & 47.254               \\
\hline
\multicolumn{10}{l}{\small {*$\hat{\mathbf{\Phi}}$/$\hat{\mathbf{Z}}$/$\hat{\mathbf{F}}$: identified mode shapes/damping ratios/natural frequencies; SD: Standard Deviation.}} \\
\end{tabular}
\end{table}

The difference between Table \ref{tab:PI_popu1_noisy} and Table \ref{tab:PI_popu1} (no noise) demonstrates that the introduction of measurement noise slightly reduces the accuracy of identifying mode shapes and damping ratios. However, the average and maximum errors for the identified natural frequencies increase more heavily. 
This discrepancy may arise because mode shapes and damping ratios are relative measures, while natural frequencies are absolute values. The addition of noise alters the absolute amplitude of the resulting PSDs without perceptibly changing the relative amplitude between the PSD of different nodes.

\subsection{Effect of training set size}
As mentioned above, the proposed population-based modal identification method necessitates data from a group of structures to train the GNN-based model. Intuitively, a larger training set provides more diversified data, which in turn enhance the performance of the trained model. While this is easier to configure when employing a simulated dataset, where populations are artificially generated, the situation is different when aiming to use real-world datasets for training. In real-world applications, the number of monitored structures that can serve for model training is typically scarce, while the number of available sensors is also limited. The latter challenge is often tackled via using of mobile or wide-coverage (such as computer vision measurements) sensor setups. In accounting for the first challenge, we here investigate the impact of the training set size on the model's performance. In this section, we use the first 40 and 200 trusses (10\% and 50\% of the original training set size) among the simply-supported population to train the model. Accordingly, the training batch size is changed to 8 and 32 to accommodate the shrinkage of the training set. The trained models are then used to identify the modal properties of 100 trusses from the testing set, with the results shown in Table \ref{tab:PI_popu1_train40} and Table \ref{tab:PI_popu1_train200}.
\begin{table}[!htp] 
\centering
\caption{Performance indicators on the testing set from the simply-supported structural population - using 40 trusses for training}
\label{tab:PI_popu1_train40}
\begin{tabular}{cccccccccc}
\hline
& \multicolumn{3}{c}{MAC of $\hat{\mathbf{\Phi}}$} & \multicolumn{3}{c}{Errors (\%) of $\hat{\mathbf{Z}}$} & \multicolumn{3}{c}{Errors (\%) of $\hat{\mathbf{F}}$} \\
& Mean   & SD  & Min   & Mean   & SD  & Max  & Mean     & SD   & Max      \\ 
\hline
Mode 1 & 0.980                & 0.028                & 0.851                & -0.409               & 0.061                & 0.603                & 1.200                & 17.916               & 84.798               \\
Mode 2 & 0.907                & 0.114                & 0.373                & 3.255                & 7.734                & 24.472               & -2.144               & 15.192               & 72.254               \\
Mode 3 & 0.796                & 0.129                & 0.280                & 2.990                & 10.015               & 27.516               & -1.869               & 12.753               & 34.363               \\
Mode 4 & 0.871                & 0.147                & 0.148                & 6.033                & 8.407                & 33.502               & -1.114               & 18.301               & 67.220               \\
\hline
\multicolumn{10}{l}{\small {*$\hat{\mathbf{\Phi}}$/$\hat{\mathbf{Z}}$/$\hat{\mathbf{F}}$: identified mode shapes/damping ratios/natural frequencies; SD: Standard Deviation.}} \\
\end{tabular}
\end{table}

\begin{table}[!htp] 
\centering
\caption{Performance indicators on the testing set from the simply-supported structural population - using 200 trusses for training}
\label{tab:PI_popu1_train200}
\begin{tabular}{cccccccccc}
\hline
& \multicolumn{3}{c}{MAC of $\hat{\mathbf{\Phi}}$} & \multicolumn{3}{c}{Errors (\%) of $\hat{\mathbf{Z}}$} & \multicolumn{3}{c}{Errors (\%) of $\hat{\mathbf{F}}$} \\
& Mean   & SD  & Min   & Mean   & SD  & Max  & Mean     & SD   & Max      \\ 
\hline
Mode 1 & 0.995                & 0.005                & 0.969                & -0.419               & 0.057                & 0.564                & 0.776                & 8.953                & 31.528               \\
Mode 2 & 0.952                & 0.074                & 0.616                & 1.073                & 5.478                & 17.137               & 0.049                & 7.560                & 31.599               \\
Mode 3 & 0.894                & 0.119                & 0.392                & 3.538                & 8.338                & 26.735               & -1.032               & 9.446                & 26.195               \\
Mode 4 & 0.897                & 0.137                & 0.283                & 0.139                & 6.725                & 22.277               & -0.795               & 11.570               & 42.272               \\
\hline
\multicolumn{10}{l}{\small {*$\hat{\mathbf{\Phi}}$/$\hat{\mathbf{Z}}$/$\hat{\mathbf{F}}$: identified mode shapes/damping ratios/natural frequencies; SD: Standard Deviation.}} \\
\end{tabular}
\end{table}

By comparing Table \ref{tab:PI_popu1_train40} (40 trusses for training), Table \ref{tab:PI_popu1_train200} (200 trusses for training), and Table \ref{tab:PI_popu1} (400 trusses for training), we find that:
\begin{itemize}
\item As anticipated, reducing the size of the training set will decrease the accuracy of modal identification. However, even with a training set that amounts to only 40 trusses (10\% of the original size), the decrease in accuracy can be deemed acceptable. A training set that amounts to 50\% of the original size can produce results similar to those obtained with the full set. Thus, while a larger training set improves modal identification, a smaller one can still be effective when more data is unavailable.
\item Compared against mode shapes, the decrease in the size of the training set induces higher errors in the identification of damping ratios and natural frequencies, especially in terms of maximum relative errors. A possible explanation to this is postulated in that mode shapes form spatial quantities, whose nature is more graph-structured than natural frequencies and damping ratios (which are global variables). Thus, the GNN can likely more robustly tackle mode shape identification, even at smaller training sets.
\end{itemize}

\subsection{Comparison against existing modal identification methods}
\label{sec:comparison}
To thoroughly evaluate our model with respect to modal identification, we compare it against one of most widely adopted modal identification methods, namely the Frequency Domain Decomposition (FDD) method (\cite{brincker2000modal,brincker2001damping}), which also operates in the frequency domain. The original FDD method requires a manual peak-picking process to determine the modes that will be identified. For the population-based modal identification, however, manually picking peaks would be too time-consuming given the high number of trusses in our testing set. To tackle this problem, we couple the FDD with an automated function (in this case the `findpeaks' function in MATLAB) to automatically pick the peaks that are corresponding to the first four structural modes. The FDD is subsequently implemented to identify mode shapes, damping ratios, and natural frequencies of the picked four modes. The MATLAB codes of our automated FDD method will be made public in a dedicated GitHub repository (\cite{Jian2024}) once this work is published. Table \ref{tab:PI_popu1_FDD} outlines the statistical indicators of FDD's modal identification performance on the testing set of the simply-supported structural population. 

\begin{table}[!htp] 
\centering
\caption{Performance indicators of FDD on the testing set from the simply-supported structural population}
\label{tab:PI_popu1_FDD}
\begin{tabular}{cccccccccc}
\hline
& \multicolumn{3}{c}{MAC of $\hat{\mathbf{\Phi}}$} & \multicolumn{3}{c}{Errors (\%) of $\hat{\mathbf{Z}}$} & \multicolumn{3}{c}{Errors (\%) of $\hat{\mathbf{F}}$} \\
& Mean   & SD  & Min   & Mean   & SD  & Max  & Mean     & SD   & Max      \\ 
\hline
Mode 1 & 1.000 & 0.000 & 1.000 & 168.034 & 61.472 & 374.388 & -0.021 & 1.004 & 2.998  \\
Mode 2 & 0.990 & 0.100 & 0.002 & 72.789 & 52.787 & 310.786 & -0.806 & 5.469 & 0.980  \\
Mode 3 & 0.987 & 0.086 & 0.174 & 0.042 & 27.217 & 96.304 & -0.824 & 2.827 & 6.636  \\
Mode 4 & 0.949 & 0.215 & 0.000 & -10.741 & 21.554 & 41.097 & -0.175 & 10.040 & 69.684  \\
\hline
\multicolumn{10}{l}{\small {*$\hat{\mathbf{\Phi}}$/$\hat{\mathbf{Z}}$/$\hat{\mathbf{F}}$: identified mode shapes/damping ratios/natural frequencies; SD: Standard Deviation.}} \\
\end{tabular}
\end{table}

The comparison between Table \ref{tab:PI_popu1_FDD} and Table \ref{tab:PI_popu1} (the GNN-based model) shows that our GNN-based model significantly outperforms the FDD in identifying damping ratios. However, when it comes to identifying mode shapes and natural frequencies, the FDD method shows substantial advantages over the GNN-based model according to the average accuracy indicators. Nevertheless, due to outliers caused by incorrectly picked peaks, the extreme errors of FDD result significantly larger when compared against those of the GNN-based model. This outcome can be attributed to the general effectiveness of FDD in identifying natural frequencies and mode shapes, coupled with its typical limitations in accurately identifying damping ratios.

Although the FDD is more accurate in general, our GNN-based model is much more efficient when performing automated modal identification for a structural population, since the FDD consumes substantial computation time to construct the PSD matrix and conduct the singular value decomposition. For example, in this numerical experiment we run both the FDD and our GNN-based model on the same CPU. The FDD spends 1096.392 seconds identifying the modal properties of 100 trusses, whereas our GNN-based model completes the same task in just 9.402 seconds. Even if we count in the training time of the GNN-based model, which is 443.838 seconds as shown in Table \ref{tab:GNNmodel}, our model remains faster than FDD. Considering the training time would, however, be an unfair comparison, since the intention (and main benefit) is to deploy the GNN-based modal identification approach in testing mode. This advantage becomes even more significant when dealing with larger structural populations, rendering the proposed GNN-based model an efficient tool for population-based modal identification.

\subsection{Effect of different structural populations}
\label{sec:population}
\begin{figure}[!htp] 
   \centering 
   \includegraphics[width=14.5cm]{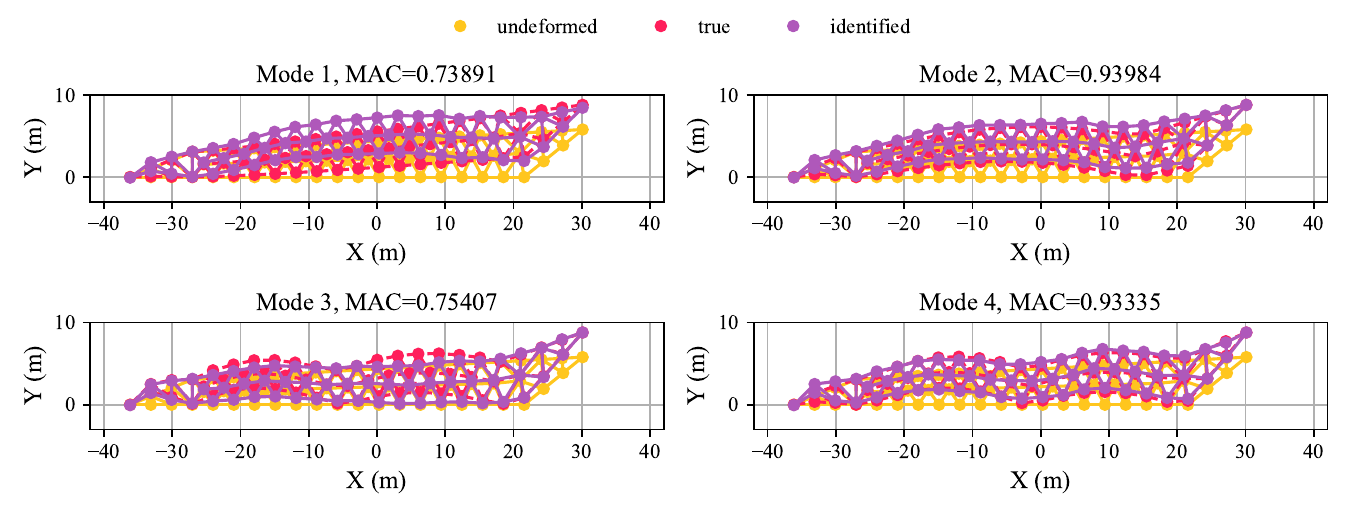} 
    \caption{Mode shape identification results of a truss from the cantilevered population}
   \label{fig:ms_popu2}
\end{figure}

\begin{figure}[!htp] 
   \centering 
   \begin{tabular}{ccc}
   \includegraphics[width=4.5cm]{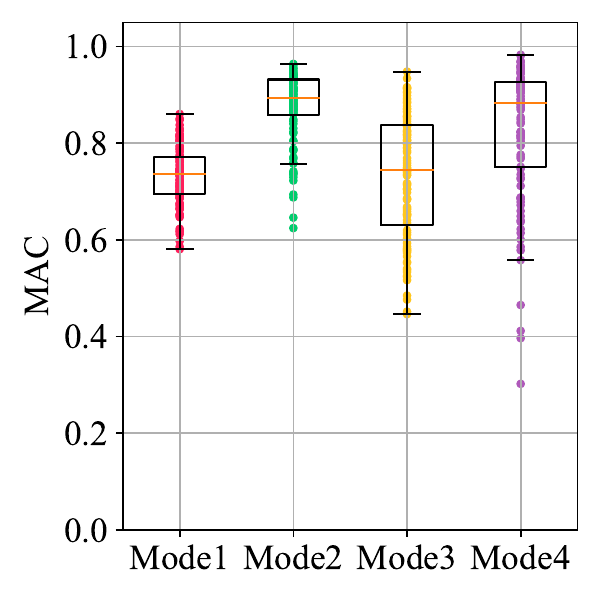} & \includegraphics[width=4.5cm]{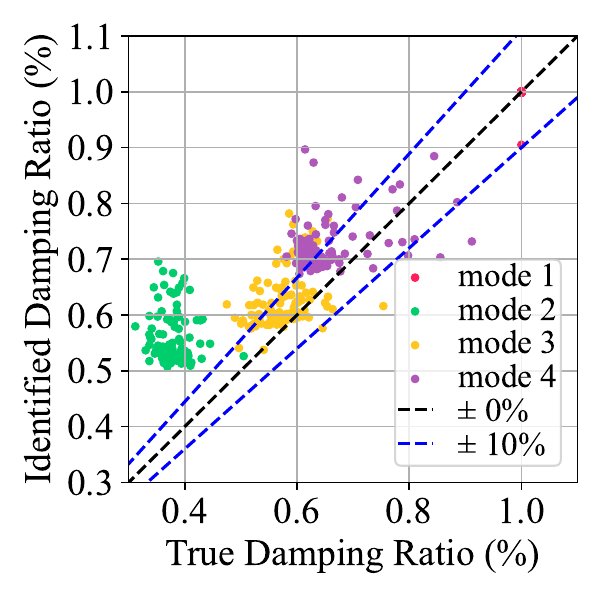} & \includegraphics[width=4.5cm]{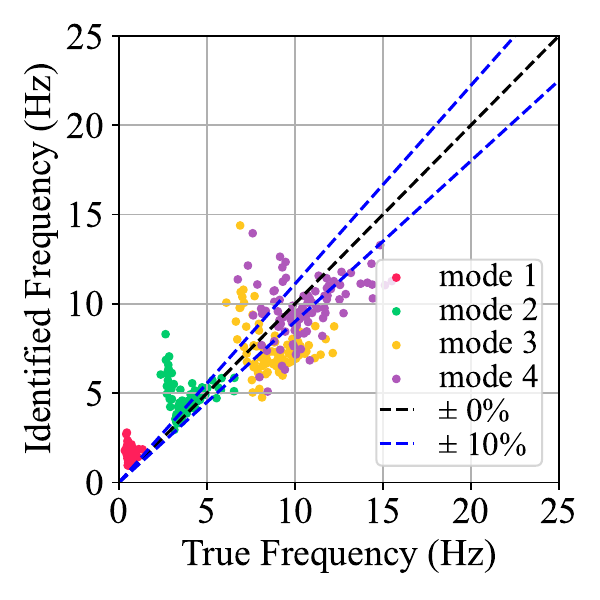} \\
   (a) & (b) & (c) \\
   \end{tabular}
    \caption{Performance of the trained model on the cantilevered truss dataset: (a) Box plot of MAC values of identified mode shapes; (b) Scatter plot of identified damping ratios; (c) Scatter plot of identified natural frequencies}
   \label{fig:test_popu2}
\end{figure}

Last but not least, we investigate the generalization capability of the proposed model on different structural populations. We apply the model trained on the simply-supported truss population, as shown in Figure \ref{fig:dataset}(a), to the cantilevered truss population, as shown in Figure \ref{fig:dataset}(b). As an example, Figure \ref{fig:ms_popu2} illustrates the identified mode shapes of a truss from the cantilevered population. The modal identification results of all (100) trusses are visualized in Figure \ref{fig:test_popu2}, where each dot represents the identification result for a certain mode of a truss. Table \ref{tab:PI_popu2} statistically summarizes the identification results illustrated in Figure \ref{fig:test_popu2}.

\begin{table}[!htp] 
\centering
\caption{Performance indicators on the testing set from the cantilevered structural population}
\label{tab:PI_popu2}
\begin{tabular}{cccccccccc}
\hline
& \multicolumn{3}{c}{MAC of $\hat{\mathbf{\Phi}}$} & \multicolumn{3}{c}{Errors (\%) of $\hat{\mathbf{Z}}$} & \multicolumn{3}{c}{Errors (\%) of $\hat{\mathbf{F}}$} \\
& Mean   & SD  & Min   & Mean   & SD  & Max  & Mean     & SD   & Max      \\ 
\hline
Mode 1 & 0.732                & 0.062                & 0.581                & -0.180               & 0.938                & 9.496                & 151.939              & 106.071              & 559.304              \\
Mode 2 & 0.874                & 0.074                & 0.625                & 48.659               & 16.558               & 97.241               & 30.832               & 44.656               & 213.919              \\
Mode 3 & 0.730                & 0.128                & 0.446                & 10.032               & 9.172                & 33.428               & -12.744              & 25.746               & 108.983              \\
Mode 4 & 0.826                & 0.141                & 0.302                & 11.352               & 9.787                & 45.934               & -5.295               & 20.704               & 83.603               \\
\hline
\multicolumn{10}{l}{\small {*$\hat{\mathbf{\Phi}}$/$\hat{\mathbf{Z}}$/$\hat{\mathbf{F}}$: identified mode shapes/damping ratios/natural frequencies; SD: Standard Deviation.}} \\
\end{tabular}
\end{table}

The identification results on the cantilevered structural population are clearly worse than those on the simply-supported structural population. This disparity is expected, given the differences between the cantilevered and simply-supported datasets, with the former not being utilized during the training of the GNN-based model. Nonetheless, the identification results are not entirely unsatisfactory. For instance, although the relative errors of identified natural frequencies and damping ratios are high, as shown in Table \ref{tab:PI_popu2}, the identified values do not deviate significantly from their target values as Figure \ref{fig:test_popu2} (b) and (c) shows. This indicates that the GNN-based model exhibits a degree of generalization capability, enabling it to leverage commonalities between two distinct yet similar structural populations.

\section{Conclusions}
To exploit the information contained within a structural population for population-based structural health monitoring, this study develops an automated operational modal analysis method capable of autonomously identifying modal properties of structures among a population by using a GNN-based deep learning model. A series of numerical experiments are carried out to validate the proposed method. Main findings emerging from this study are summarized as follows:

1) \textit{Model architecture}. The model comparison suggests that the GraphSAGE model outperforms the GCN and GAT model. The ablation study reveals that using an MLP to encode the PSD input of the model is beneficial. More importantly, the non-GNN model is inferior to all the three GNN-based models, confirming the superiority of GNNs in processing graph-structured data.

2) \textit{Modal identification performance}. Upon training on data from a structural population built using the finite element method and model updating, the proposed GNN-based model can efficiently and effectively identify natural frequencies, damping ratios, and mode shapes of various structures in the same structural population. The performance is robust even in the presence of measurement noise and spatially sparse measurements. To tackle the issue of measurement sparsity, the feature propagation algorithm is needed to reconstruct the full-field node input features, specifically the PSD of node acceleration, before using the proposed model to process incomplete measurements.

3) \textit{Limitations}. While the proposed model is significantly faster than the classic frequency domain decomposition method and provides more accurate damping ratio identification, it is less accurate in identifying natural frequencies and mode shapes. Furthermore, when training data is insufficient, or the testing structure belongs to a different population, the model accuracy decreases, albeit still considered acceptable. This is expected due to the purely data-driven nature of the proposed approach. Future research could incorporate some physical information to enhance the model's generalization ability. A further limitation is that the model can only identify absolute mode shapes, instead of signed mode shapes, owing to the use of the auto-PSD that lacks phase information. While this is not prohibitive for SHM purposes, additional node features besides PSD could be utilized to address this issue. Lastly, it would be valuable to validate the proposed method with real-world data.

\begin{Backmatter}

\paragraph{Funding Statement}
The research was conducted at the Singapore-ETH Centre, which was established collaboratively between ETH Zurich and the National Research Foundation Singapore. This research is supported by the National Research Foundation, Prime Minister’s Office, Singapore under its Campus for Research Excellence and Technological Enterprise (CREATE) program.

\paragraph{Competing Interests}
The authors declare no competing interests exist.

\paragraph{Data Availability Statement}
Demonstrative MATLAB and Python codes that implement the proposed method are openly available at
\href{https://github.com/JxdEngineer}{https://github.com/JxdEngineer}.

\paragraph{Ethical Standards}
The research meets all ethical guidelines, including adherence to the legal requirements of the study country.

\paragraph{Author Contributions}
Conceptualization: Eleni Chatzi; Xudong Jian. Methodology: Xudong Jian; Yutong Xia; Gregory Duthé; Kiran Bacsa; Wei Liu. Data curation: Xudong Jian. Formal analysis: Xudong Jian. Software: Yutong Xia. Data visualisation: Xudong Jian. Writing - original draft: Xudong Jian. Writing - Review \& Editing: Eleni Chatzi; Xudong Jian; Kiran Bacsa; Yutong Xia; Gregory Duthé. Supervision: Eleni Chatzi. Funding Acquisition: Eleni Chatzi. Project administration: Eleni Chatzi. All authors approved the final submitted draft.


\printbibliography

\end{Backmatter}

\end{document}